\documentclass{caosp}

\usepackage{graphicx}

\articleNo{413/8}
\pubyear{2011}
\volume{35}
\volnumber{3}
\firstpage{1}
\received{January 3, 2011}
\accepted{April 7, 2011}

\begin{document}

\hauthor{Ph.-A.\,Bourdin}

\title{Denoising observational data}

\author{Ph.-A.\,Bourdin \inst{1,} \inst{2}}

\institute{
           Max-Planck-Intitute for Solar System Research,\\
           Max-Planck-Str. 2
           D-37191 Katlenburg-Lindau, Germany \\ \email{Bourdin@MPS.mpg.de}
         \and 
           Kiepenheuer-Institut f\"ur Sonnenphysik,\\
           Sch\"oneckstr. 6,\\
           D-79104 Freiburg, Germany 
          }

\date{January 3, 2011}

\maketitle

\begin{abstract}
  Reducing noise caused by the instrumentation in observational data
  is a crucial step in data post-processing.  A method is searched
  for that conserves most of the instrumental resolution and
  introduces as few methodical artefacts as possible.  With such a
  method integrated in an observation sites software tool-chain, the
  resources spent for the generation of observational data will more
  likely find their way into resulting scientific publications;
  otherwise, for data post-processing often methods are used, which
  just smear out the noise, introduce artefacts, or decrease the
  provided resolution in space or time.  A short review of
  different techniques is given here, and a non-local averaging method
  is applied to Hinode magnetograms and G-band data.  The presented
  method fits the needs for various kinds of observational data.
\end{abstract}

\section{Introduction}

Any modern telescope or observational instrument is nowadays equipped
with digital sensors, in particular CCDs for imaging.  A resolution
element of a CCD basically counts photons and provides 2D pixel data
with some kind of noise on each pixel.  The cause for the noise might
lie in the CCD or other parts of the instrument, but usually the
noise can be treated as it would have come from only one source.  Good
CCD cameras often use 16 bit unsigned integers and can reach photon
counts up to $2^{16}-1 = 65535$.  Typical signal-to-noise ratios
combined with reasonable exposure times of such CCDs are usually
resulting in a noise count of 10 or more, where a Gaussian
distribution of the photon noise can be assumed.  Therefore denoising
methods can be tested, by adding Gaussian noise to a known image, run
the denoising method, and compare the resulting images with the
original ones.  Presented here are excerpts from a well written review
article (Buades {\it et al.}, 2006) as well as an analysis of a
non-local averaging method (Buades {\it et al.}, 2005) applied to
solar observations, such as Stokes-V images from the Hinode satellite
as well as G-band images from the Dutch Open Telescope (DOT).

\section{Common methods for denoising}

As a young scientist or PhD student, who has to post-process
observational data for the first time, the obvious methods to get rid
of noise are smoothing the data in space, time-averaging aligned image
data, or simply reducing the resolution to the maximum possible, so that
the noise cannot be seen.  All of those methods basically mean not
removing any noise, but smearing it out in space or time.  Another
possibility is to try a Gaussian convolution method, which tries to
locally fit the image by Gaussian curves.  This method also smooths an
image, lowers the resolution, and cannot maintain sharp contrasts, as
it can be seen in difference plots of a noisy and its denoised image.
More advanced methods are needed to make better use of the provided
resolution, cadence or other kinds of data qualities.

In principle, denoising methods can be divided into local and non-local
methods.  Where averaging methods are now available in local and
non-local variants, any fitting method is local and any frequency
domain (Fourier transform based) method is non-local.  The advantage
of non-local methods is clearly the possibility to profit from
self-similarities of any scale, where local methods have the dilemma
of delivering performance by using larger amounts of pixels versus
maintaining resolution by using less amounts of pixels.

Fitting methods ({\it e.g.}, wavelets) use subfields of an image to fit a set
of base functions to the shape of the image.  Because such sets are
discrete and finite, the fit is usually not representing the true
shape of the image and it introduces artefacts at the borders of the
subfields.  Total variation (TV) minimization, iterated TV,
anisotropic filtering, and entropy reduction methods are usually
producing an oil-painting effect or deliver low denoising performance.

Frequency domain methods ({\it e.g.}, Fourier-Wiener and DCT-empirical Wiener
filters) are introducing artefacts, which are then uniformly visible
{\it e.g.}, as wiggles in large parts of the denoised image.  Furthermore, no
clear distinction can be made between high frequencies resulting from
noise or from true sharp contrasts on a low spatial scale.  Some
smoothing can therefore not be avoided.  A more in-depth discussion of
all these methods can be found in the mentioned review article (Buades
{\it et al.}, 2006).

\section{A non-local averaging method}

The NL-means algorithm features the advantages of a non-local method,
without introducing artefacts.  This is achieved by building, to get
one pixel of the denoised image, a superposition (or an average) of
all pixels in the image, weighted by the similarity between the
surrounding area of the to-be-denoised pixel and the surrounding area
of every other pixel in the whole image.  Where there is a strong
similarity, the contribution to the average is strong; where there is
no similarity, the contribution is negligible.  Furthermore, a small
fixed fraction of the original image is added to the result, to
maintain the original image in areas, where there is less similarity to
other areas.  This conserves unique features in a noisy image, if they
are significantly above the noise level (like {\it e.g.}, a G-band
bright feature).  A full description of the algorithm can be found in
(Buades {\it et al.}, 2005).

For better denoising performance, the surrounding areas around pixels
(windows) can be varied in size and shape.  A smooth window shape can
be achieved by using a quadratic window with a Gaussian kernel
multiplied to it.  The half-width of the Gaussian kernel should match
at least two optical resolution elements, which can be more than two
pixels, depending on the instrument.  For faster computation the
window size should be kept small, since every possible relation
between two pixels of the image needs to be computed.  Nonetheless,
using parallel programming techniques, it is possible to denoise a
1\,k$\times$1\,k pixels image within a few seconds on an off-the-shelf
server hardware with 8 CPU cores.  The computation domain for one CPU
core should be sized such that all necessary image data fits into the
L2-cache; for larger images, more CPU cores are needed.  A special
feature of the NL-means algorithm is the fact that it can easily be
applied also to time series of images without previous alignment,
since the algorithm is by itself looking for self-similarities and can
use windows from other frames of the time series for averaging.  This
is described in the movie denoising article of the same authors
(Buades {\it et al.}, 2008).

\section{Application to solar observations}

\begin{figure}
 \centerline{\includegraphics[width=11.5cm]{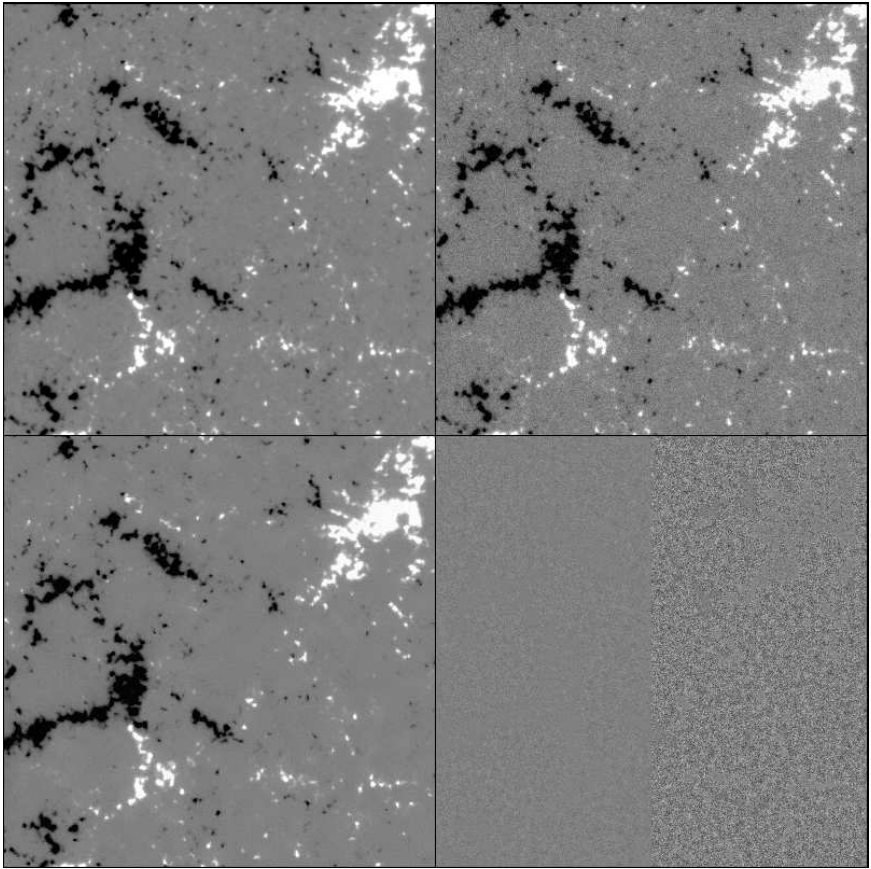}}
\caption{A Hinode/NFI Stokes-V map of a small active region.  The peak
  polarities have values of around 1200 Gauss, in this image the
  saturation level is at 300 Gauss and corresponds to a pixel value of
  127.  The upper left pane shows the original image, in the upper
  right pane a Gaussian noise with a standard deviation of 10 was added.
  The lower left pane shows the denoised image as denoised with the
  NL-means algorithm.  The lower right pane shows the difference of
  the noisy and the denoised image; in the right half the contrast has
  been improved by a factor of 3 for better visibility.}
\label{f1}
\end{figure}

Fig.\,\ref{f1} shows a Hinode/NFI Stokes-V map of a small active
region (pixel value range is -128 to 127), which has been used to test
NL-means by adding Gaussian noise with a standard deviation of 10.
This corresponds to a high noise level, compared to the original
Hinode data.  The denoising result shows most of the features of the
original image, except some small-scale low-signal features, which are
anyway not above the given noise level.  The difference plot shows
that very little of the actual structure of the original image has been
falsely recognized as noise.  The astounding result is that the
algorithm worked that well because one cannot say that the original
image would present much self-similarity, but nonetheless quiet Sun
areas between the strong polarities have been well denoised.
Furthermore, one could say that there is a loss of "visible by
eye"-features in the quiet Sun area, but one has to notice that these
features might be recognized by our brain as "visible", but may not be
mathematically significant, because they are below the noise level.
So, one could see this effect as a benefit instead of a defect,
because by applying this denoising method we give our selective
recognition a simple proxy for the significance of certain features.

\begin{figure}
\centerline{\includegraphics[width=1.025\textwidth]{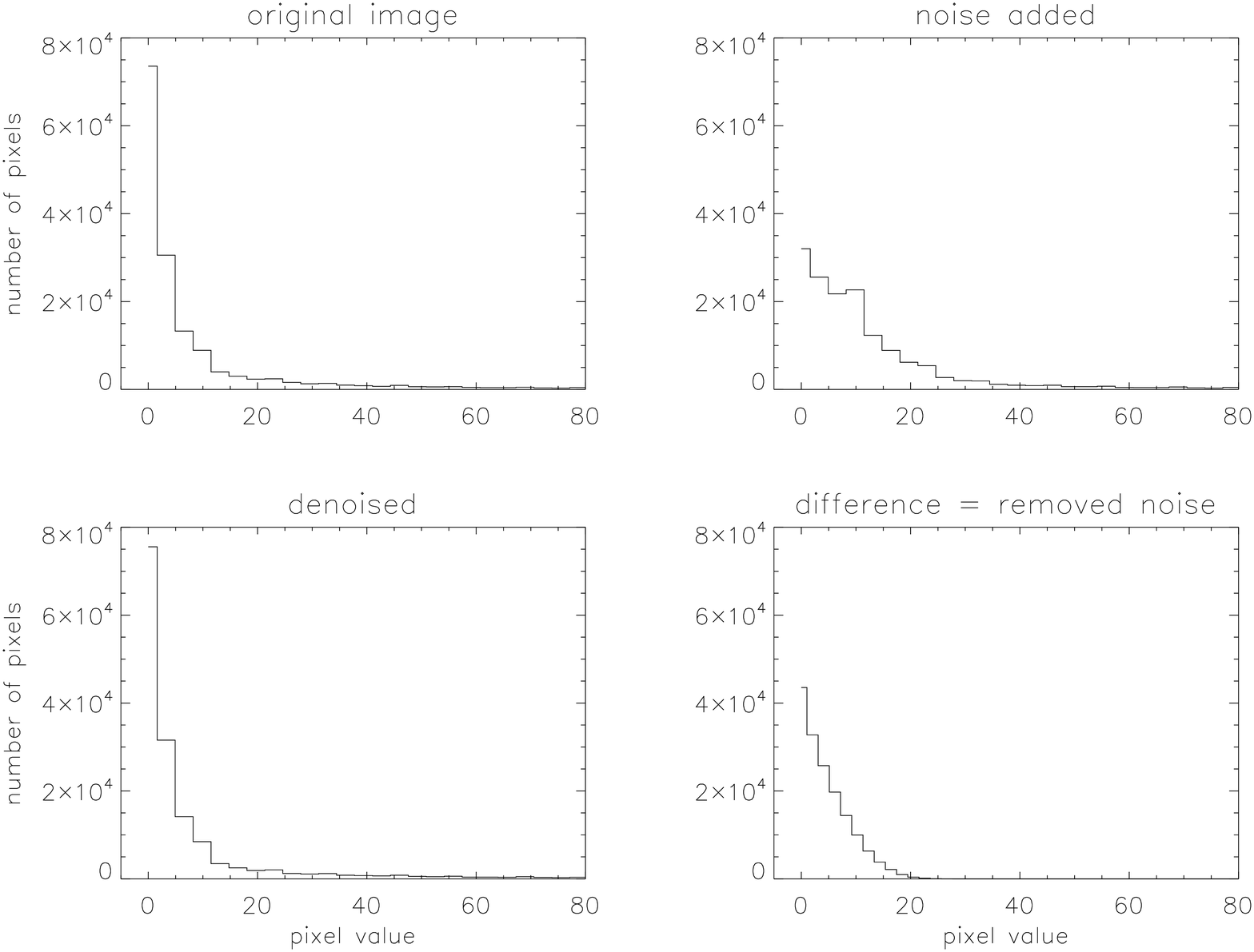}}
\caption{Histograms of corresponding images shown in Fig.\,\ref{f1} in
  the same ordering.}
\label{f2}
\end{figure}

A remarkable result is shown in the histograms in Fig.\,\ref{f2},
where one can see again the original, noisy, denoised, and difference
images in the same order as in Fig.\,\ref{f1}.  Even though the
algorithm only knows the noisy image with a relatively flat histogram,
the method is capable of recovering an image that has a histogram very
close to the original one.  In the histogram of the removed noise one
can recognize its standard deviation.

\begin{table}
\small
\begin{center}
  \caption{Bias of the NL-means denoising method for different noise
    levels.}
\label{t1}
\begin{tabular}{cc}
\hline\hline
noise level  &  bias \\
\hline
0   & 0.008\\
5   & 0.009\\
10  & 0.016\\
15  & 0.032\\
20  & 0.029\\
\hline\hline
\end{tabular}
\end{center}
\end{table}

\begin{table}
\small
\begin{center}
  \caption{Standard deviation and mean value of the images in
    Fig.\,\ref{f1}.}
\label{t2}
\begin{tabular}{ccc}
\hline\hline
image  &  standard deviation  &  mean value \\
\hline
original    & 32.9  & 2.061\\
noisy       & 33.7  & 2.060\\
denoised    & 32.0  & 2.044\\
difference  & 7.2  & 0.016\\
\hline\hline
\end{tabular}
\end{center}
\end{table}

Tab.\,\ref{t1} shows the bias of the denoising algorithm for different
noise levels.  The mean value of the original image is 2.06 and the
standard-deviation is 32.9 as it can be seen in Tab.\,\ref{t2}, where
the standard deviation and mean value are given for the noise level 10
images, too.  It is found that the Bias in that case is around 0.8\,\%
of the mean value and below 1.6\,\% at noise levels of 15 and
20. Levels above 10 are already much higher than the noise level one
would expect in the provided original image data.

\begin{figure}[t]
\centerline{\includegraphics[width=11.5cm]{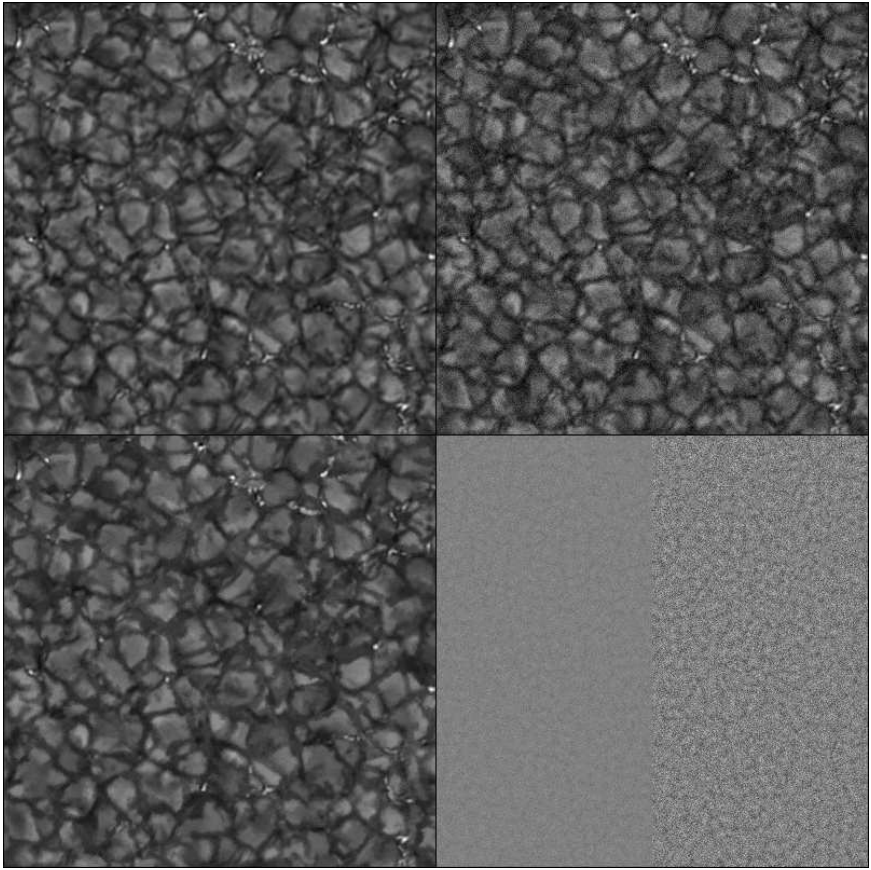}}
\caption{A G-band image of solar granulation and bright features as
  taken from the Dutch Open Telescope (DOT) after
  speckle reconstruction.  The panels are in the same order as in
  Fig.\,\ref{f1}, a Gaussian noise of level 10 has been used to test
  the NL-means denoising method.}
\label{f3}
\end{figure}

Fig.\,\ref{f3} shows the same method applied to a G-band image with a
noise of level 10 added to it.  Basically, the same findings as above
are seen.  In the inter-granular lanes we see the biggest differences
to the original images, because some of the original contrast is
hidden in the noise level and gets flattened out.  One should also
notice the conservation of G-band bright features without introduction
of artefacts and without any loss of spatial resolution.  With a more
realistic (lower) noise level and with 16-bit data, better results are
achievable.

\section{Conclusion}

NL-means should be considered as denoising tool for solar
observational data.  Losses of image features during denoising are
usually only because the features are anyway not above the level of
significance.  Where there is no possibility to improve image quality,
NL-means would also not do much harm.  Integration of such a denoising
method in the observation sites software tool-chain could improve the
throughput of valuable image quality from the detector into the
scientific publication.

\acknowledgements 
Hinode is a Japanese mission developed and launched by ISAS/JAXA, with
NAOJ as domestic partner and NASA and STFC (UK) as international
partners. It is operated by these agencies in co-operation with ESA
and NSC (Norway). The DOT was built by the Sterrekundig Instituut
Utrecht, the Physics Instrumentation Group of Utrecht University, and
the Central Workshop of Delft Technical University.  The DOT
completion, installation and verification are funded by the Stichting
Technische Wetenschappen (STW) of the Netherlands Organization for
Scientific Research (NWO).  The current installation at La Palma
proceeds under the umbrella of the Instituto de Astrof\'isica de
Canarias.

\end{document}